\documentclass[a4paper,11pt]{article}
\pdfoutput=1

\usepackage{jheppub} 

\usepackage{amsmath,amssymb,bbold,slashed}
\usepackage{graphicx, soul}
\usepackage{tabu}
\usepackage[usenames,dvipsnames]{xcolor}

\title{Gaugino Mediation Scenarios for Muon $g-2$ and Dark Matter}

\author[a]{Peter Cox}
\author[a]{Chengcheng Han}
\author[a,b,c]{Tsutomu T. Yanagida}
\author[d]{Norimi Yokozaki}

\affiliation[a]{Kavli Institute for the Physics and Mathematics of the Universe (WPI), The University of Tokyo Institutes for Advanced Study, University of Tokyo, Kashiwa 277-8583, Japan}
\affiliation[b]{T. D. Lee Institute and School of Physics and Astronomy, Shanghai Jiao Tong University, Shanghai 200240, China}
\affiliation[c]{Hamamatsu Professor}
\affiliation[d]{Department of Physics, Tohoku University, Sendai, Miyagi 980-8578, Japan}

\emailAdd{peter.cox@ipmu.jp}
\emailAdd{chengcheng.han@ipmu.jp}
\emailAdd{tsutomu.tyanagida@ipmu.jp}
\emailAdd{yokozaki@tuhep.phys.tohoku.ac.jp}

\preprint{IPMU18-0193, TU-1077}

\abstract{
We explore the possibility that the muon $g-2$ anomaly and the nature of dark matter can be simultaneously explained within the framework of gaugino mediation, focusing on bino-like dark matter where the observed abundance is obtained via co-annihilations. 
The minimal model with non-universal gaugino masses is excluded by stau vacuum instability, although this constraint can be somewhat relaxed via the addition of a universal soft scalar mass (or $B-L$ gaugino mediation).  
A more promising alternative is gaugino+Higgs mediation, which significantly raises the soft masses of the third generation sfermions leading to a split spectrum. 
In this framework, the muon $g-2$ can be easily explained and the dark matter abundance obtained through either bino-wino or bino-slepton co-annihilations.
}

\notoc

\begin{document} 

\maketitle
\flushbottom
\newpage

%%%%%%%%%%%%%%%%%%%%%%%%%%%%%%%%%%

\section{Introduction}

Supersymmetry (SUSY) remains an attractive and well-motivated framework for physics beyond the Standard Model. 
Nevertheless, its absence in direct searches at the LHC suggests that the spectrum of coloured superpartners should be somewhat heavier than our expectations based on naturalness. 
On the other hand, the long-standing discrepancy in the anomalous magnetic moment of the muon~\cite{hep-ex/0602035} can be straightforwardly explained by the presence of weak-scale sleptons and gauginos, suggesting that SUSY partners in the electroweak sector might be somewhat lighter. 
This possibility will be brought into sharper focus in the near future, with new experiments that will measure the muon $g-2$ with improved precision~\cite{1701.02807}. 

Furthermore, the electroweak particles involved in explaining the muon $g-2$ could also play a role in producing the observed dark matter density; in this case the lightest neutralino provides the dark matter candidate, with its stability guaranteed by R-parity.\footnote{The muon $g-2$ can also be explained in the context of gauge mediation where the gravitino is the lightest supersymmetric particle (LSP)~\cite{1805.01607}.} 
This interesting possibility has attracted significant attention, with many studies conducted in the context of the Minimal Supersymmetric Standard Model (MSSM)~\cite{1404.4841, 1406.6925, 1409.3930, 1503.08219, 1504.00505, 1505.05877, 1505.05896, 1507.01395, 1608.03641, 1704.05287, 1710.11091}. 
However, such a scenario already runs into significant tension for a generic well-tempered neutralino, with the stringent bounds from dark matter direct detection forcing one into tuned regions of parameter space such as the blind-spot~\cite{1211.4873, 1404.0392, 1612.02387, 1701.02737} and funnel regions~\cite{1303.3040, hep-ph/0106275}. 
An alternative way to evade the direct detection constraints is to consider a bino-like LSP; the correct relic abundance can then be obtained via bino-wino~\cite{hep-ph/0511034} or bino-slepton~\cite{hep-ph/9905481} co-annihilations. 
This possibility was recently explored in detail within the low-scale MSSM~\cite{1805.02802}. 

In this work, we demonstrate how the co-annihilation scenarios identified in refs.~\cite{1710.11091, 1805.02802} for simultaneously explaining the muon $g-2$ and the dark matter abundance can be obtained within a realistic supersymmetry breaking setup. 
A significant consideration in any ultraviolet (UV) framework must be the SUSY flavour problem, since the presence of weak-scale sleptons (and squarks) will generically have disastrous consequences in flavour observables, such as $\mu\to e\gamma$.  
With this in mind, we work within the framework of gaugino mediation~\cite{Inoue:1991rk, hep-ph/9911293, hep-ph/9911323}, which provides an elegant solution to the SUSY flavour problem: the soft masses of the squarks and sleptons are initially negligible at the grand unified theory (GUT) scale and are then radiatively generated via flavour-blind gaugino loops. 

The possibility of explaining the muon $g-2$ ---but not dark matter--- in the context of gaugino mediation was previously explored in ref.~\cite{1501.07447}, where non-universal masses for the gauginos were required because of the Higgs boson mass and LHC constraints.
We find that the minimal model with only non-universal gaugino masses at the GUT scale cannot provide the observed dark matter abundance with a bino-like LSP, in addition to explaining the muon $g-2$, as it encounters problems with stau instability. 
We therefore consider two simple extensions to this minimal setup. 
Firstly, in sec.~\ref{sec:m0}, we allow for the presence of universal soft scalar masses at the GUT scale; a similar situation is obtained with pure gaugino mediation if $B-L$ is also gauged. 
We find that while there are regions of parameter space that can explain both the dark matter abundance and the muon $g-2$, these are still significantly restricted by stau instability. 
Furthermore, the best-fit region is already being tested by LHC searches for the light sleptons or chargino. 

Next, the strong constraint from stau instability motivates us to consider, in sec.~\ref{sec:Higgs-med}, a model with gaugino+Higgs mediation, where there is a direct coupling of the Higgs doublets to the SUSY breaking field. 
Such a coupling is expected if the sequestered K\"ahler potential originates from an extra dimension and the Higgs doublets live in the bulk~\cite{hep-ph/9911323}. 
In this setup, the soft SUSY breaking masses for the Higgs doublets are assumed to be tachyonic, allowing hierarchical sfermion masses to be generated from Higgs loops without inducing too large FCNC~\cite{Yin:2016shg,Yanagida:2016kag,Yanagida:2018eho}. 
The third generation sfermions are then much heavier than the first/second generation, evading the constraint from stau vacuum stability, and also helping to raise the Higgs mass to the observed value. 
In this setup, we find that there exist viable regions of parameter space that can explain the muon $g-2$ and produce the observed dark matter abundance via either bino-wino or bino-slepton co-annihilations. 
Some of this interesting parameter space will also be probed by direct searches at the (HL-)LHC in the future. 

%%%%%%%%%%%%%%%%%%%%%%%%%%%%%%%%%%

\section{Gaugino mediation with universal soft scalar masses} \label{sec:m0}

\subsection{Model} \label{sec:m0-model}

We begin by considering the following form for the K\"ahler potential and superpotential:
\begin{align} \label{eq:lag1}
  K &= -3 M_P^2 \ln \left[1 - \frac{f(Z, Z^\dag) + Q_I^\dag Q_I + \Delta K}{3M_P^2}\right],\, \notag \\
  W &= w(Z)  + \mu H_u H_d + W_{\rm Yukawas} \,, 
\end{align}
where $Q_I$ is a MSSM superfield, and $Z$ is a SUSY breaking field. 
For $\Delta K=0$, $K$ is identified as the sequestered K\"ahler potential~\cite{Inoue:1991rk, hep-th/9810155}, and the soft SUSY breaking masses for the sfermions and Higgs doublets vanish at tree-level. We assume $\left<Z\right>/M_P \ll 1$. 

The deviation from the sequestered form, $\Delta K$, is assumed to be 
\begin{equation} \label{eq:deltaK}
  \Delta K =  - d  \frac{Z^\dag Z}{M_P^2}  Q_I^\dag Q_I - c_b \frac{Z^\dag Z}{M_P^2} H_u H_d + h.c. \,.
\end{equation}
Then, the sfermions and Higgs doublets obtain universal soft masses:
\begin{equation}
  m_{Q_I}^2 = m_0^2 = d \frac{|F_Z|^2}{M_P^2} = 3 d  \left(\frac{\partial^2 K}{\partial Z \partial Z^\dag}\right)^{-1} m_{3/2}^2 \,,
\end{equation}
and the Higgs B-term arises as
\begin{align} \label{eq:higgs_b-term}
  V &= \left[ c_b \frac{|F_Z|^2 }{M_P^2} - e^{K/(2M_P^2)} \mu \frac{\left<w^*\right>}{M_P^2} \right] H_u H_d + h.c. \notag \\
  &= c_b' m_{3/2} \mu H_u H_d + h.c. \,.
\end{align}
For simplicity, we have neglected possible terms linear in $Z$ in eq.~\eqref{eq:deltaK}, which results in vanishing trilinear couplings, $A_{ijk}$. 
One could interpret the flavour-independent soft scalar masses, $m_{0}^2$, as being generated via radiative corrections from the gaugino of the $B-L$ gauge group, which is a natural extension of the MSSM gauge group (see appendix~\ref{app:B-L}). 
In this case, the Higgs soft masses vanish and the squark masses are smaller than the slepton masses at the high energy scale. 
However, these differences are not expected to lead to significant deviations in the phenomenology we will discuss in the next section.
Therefore, we will continue to use a universal soft scalar mass, $m_0$, for all the MSSM scalars. 

Next, consider the gaugino masses. 
These are required to be non-universal due to the LHC constraints and Higgs boson mass of 125\,GeV, which both require a large gluino mass; on the other hand, the bino (and sleptons) should remain light in order to explain the muon $g-2$.
The non-universal gaugino masses can naturally arise in a product group unification model, which also elegantly solves the doublet-triplet splitting problem in GUT models~\cite{hep-ph/9409329, hep-ph/9509201, hep-ph/9607463}.\footnote{For a discussion of the muon $g-2$ in models where non-universal gaugino masses arise from a SUSY breaking field $Z$ in a non-trivial SU(5) representation see ref.~\cite{1303.5830}.}
Consider, as an example, the $SU(5) \times SU(3)_H \times U(1)_{H}$ product group unification model. The Lagrangian for the field strength superfields is 
\begin{align} \label{eq:pgu_gauginos}
  \mathcal{L} &=  \int d^2\theta \left( \frac{1}{4 g_5^2}  -\frac{k_5}{2M_P} Z \right) W_5 W_5 + h.c. \notag \\
  &+\int d^2\theta \left(  \frac{1}{4 g_{3H}^2}  -\frac{k_{3H}}{2M_P} Z \right) W_{3H} W_{3H}+ h.c. \notag \\
  &+  \int d^2\theta \left( \frac{1}{4 g_{1H}^2} - \frac{k_{1H}}{2M_P} Z \right) W_{1H} W_{1H} + h.c. \,,
\end{align}
where $W_5$, $W_{3H}$ and $W_{1H}$ are the field strength superfields for $SU(5)$, $SU(3)_H$ and $U(1)_H$, respectively. 
The coefficients, $k_5$, $k_{3H}$ and $k_{1H}$, are assumed to be real.\footnote{This fact may be justified when eq.~\eqref{eq:pgu_gauginos} arises from anomalies of a symmetry under which $Z$ is charged.}
After the $SU(5) \times SU(3)_H \times U(1)_H$ is broken down to the SM gauge group, non-universal gaugino masses arise at the GUT scale:
\begin{align}
  M_1 &= (k_5\mathcal{N} + k_{1H}) \frac{g_5^2 g_{1H}^2}{g_5^2 + \mathcal{N}g_{1H}^2} \frac{F_Z}{M_P} \,, \notag \\
  M_2 &= k_5 g_5^2 \frac{F_Z}{M_P} \,, \\
  M_3 &= (k_5 + k_{3H}) \frac{g_5^2 g_{3H}^2}{g_5^2 + g_{3H}^2} \frac{F_Z}{M_P} \notag \,,
\end{align}
where the coefficient $\mathcal{N}$ depends on the $U(1)_H$ charge of the GUT breaking Higgs field. 
Note that approximate gauge coupling unification is realised in this model when $g_{3H}$ and $g_{1H}$ are sufficiently strong compared to $g_5$:
\begin{align}
  g_1^{-2} &= g_5^{-2}  + \mathcal{N}^{-1} g_{1H}^{-2} \,, \notag \\
  g_2^{-2} &= g_5^{-2} \,, \\
  g_3^{-2} &= g_5^{-2} + g_{3H}^{-2} \notag \,,
\end{align}
at the GUT scale.

Finally, let us briefly comment on CP violation. 
If CP is violated in the SUSY breaking sector, this will generically induce a large electric dipole moment of the electron, in conflict with experiment. 
To avoid this problem, we simply assume that CP is conserved in the SUSY breaking sector.\footnote{Another possible solution is to consider a shift symmetry of the SUSY breaking field: $Z \to Z + i\mathcal{R}$, where $\mathcal{R}$ is a real constant~\cite{Iwamoto:2014ywa}. In this case, only $W(Z)={\rm const}$ is allowed. Still, SUSY can be broken with a specific form of the K\"ahler potential~\cite{Izawa:2010ym}.}

%%%%%%%%%%%%%%%%%%%%%%%%%%%%%%%%%%

\subsection{Phenomenology}

\subsubsection{Muon $g-2$}

\begin{figure}[t]
  \centering
  \includegraphics[width=0.4\textwidth]{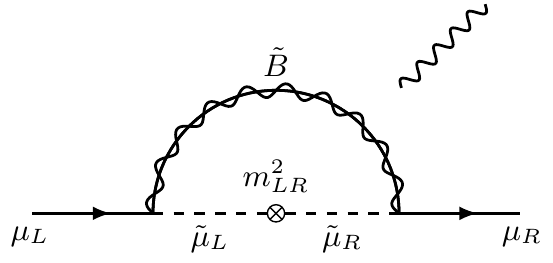}
  \caption{Bino-smuon contribution to the muon $g-2$.}
  \label{fig:g-2}
\end{figure}

For both of the models considered in this paper, the relevant regions of parameter space feature a relatively large value for $\mu$. 
The Higgsinos are therefore somewhat decoupled and do not contribute significantly to the muon $g-2$. 
The dominant SUSY contribution then comes from the bino-smuon diagram in fig.~\ref{fig:g-2}, which is proportional to the smuon mixing and hence is enhanced at large $\mu\tan\beta$. This contribution is given by~\cite{hep-ph/0103067}
\begin{equation} \label{eq:g-2}
  a_\mu^{\text{SUSY-1L}} = \frac{g_1^2}{48\pi^2} \frac{m_\mu^2 M_1 \mu\tan\beta}{m_{\tilde{\mu}_2}^2-m_{\tilde{\mu}_1}^2} \left( \frac{F_2^N(x_{1})}{m_{\tilde{\mu}_1}^2} - \frac{F_2^N(x_{2})}{m_{\tilde{\mu}_2}^2} \right) \,,
\end{equation}
where $x_{i}=M_1^2/m_{\tilde{\mu}_i}^2$, and
\begin{equation}
  F_2^N(x) = \frac{3}{(1-x)^3}\left(1-x^2+2x\ln x\right) \,.
\end{equation}
We also include the leading two-loop effects, resulting in\footnote{In our numerical results we use the complete 1-loop result, not only the contribution in eq.~\eqref{eq:g-2}.}
\begin{equation} \label{eq:Delta}
  a_\mu^\text{SUSY} = \left(1-\frac{4\alpha}{\pi}\ln\frac{m_{\tilde{\mu}}}{m_\mu}\right) \left(\frac{1}{1+\Delta_\mu}\right) a_\mu^\text{SUSY-1L} \,,
\end{equation}
where the first term in brackets corresponds to the effect of QED running down to the muon mass scale and leads to a reduction of around 10\%~\cite{hep-ph/9803384}. 
The second factor comes from resummed $\tan\beta$-enhanced corrections to the muon Yukawa coupling~\cite{0808.1530, hep-ph/9912516}; the expression for $\Delta_\mu$ can be found in ref.~\cite{0808.1530} and is proportional to $\mu\tan\beta$. 
In our analysis we use the latest result for the SM hadronic vacuum polarisation contributions~\cite{1802.02995} which gives
\begin{equation}
  a_\mu^\text{exp} - a_\mu^\text{SM} = \left(2.71 \pm 0.63 \pm 0.36\right) \times 10^{-9} \,,
\end{equation}
where the errors are from experiment and theory respectively. 
This corresponds to a $3.7\sigma$ deviation from the SM prediction. 

\subsubsection{Results}

Following the discussion of sec~\ref{sec:m0-model}, we have the following six input parameters at the GUT scale:
\begin{equation}
  M_1, M_2, M_3, m_0, \mu, B\mu \,,
\end{equation}
where $m_0$ is the universal soft scalar mass, and $|\mu|$ and $B\mu$ are substituted with $\tan\beta$ after electroweak symmetry breaking (EWSB), leaving us with five free parameters.
We have performed a scan over this parameter space in order to determine the regions that are consistent with both the measured value of the muon $g-2$ and the observed dark matter density, as well as satisfying any bounds from LHC searches. 
We used a modified version of the spectrum generator {\tt SuSpect-2.43}~\cite{hep-ph/0211331}. 
The dark matter relic density was computed using {\tt MicrOMEGAs-5.0.2}~\cite{1305.0237}, and the Higgs mass with {\tt FeynHiggs-2.14.1}~\cite{1608.01880,1312.4937}. 

\begin{figure}[t]
  \centering
  \includegraphics[width=0.6\textwidth]{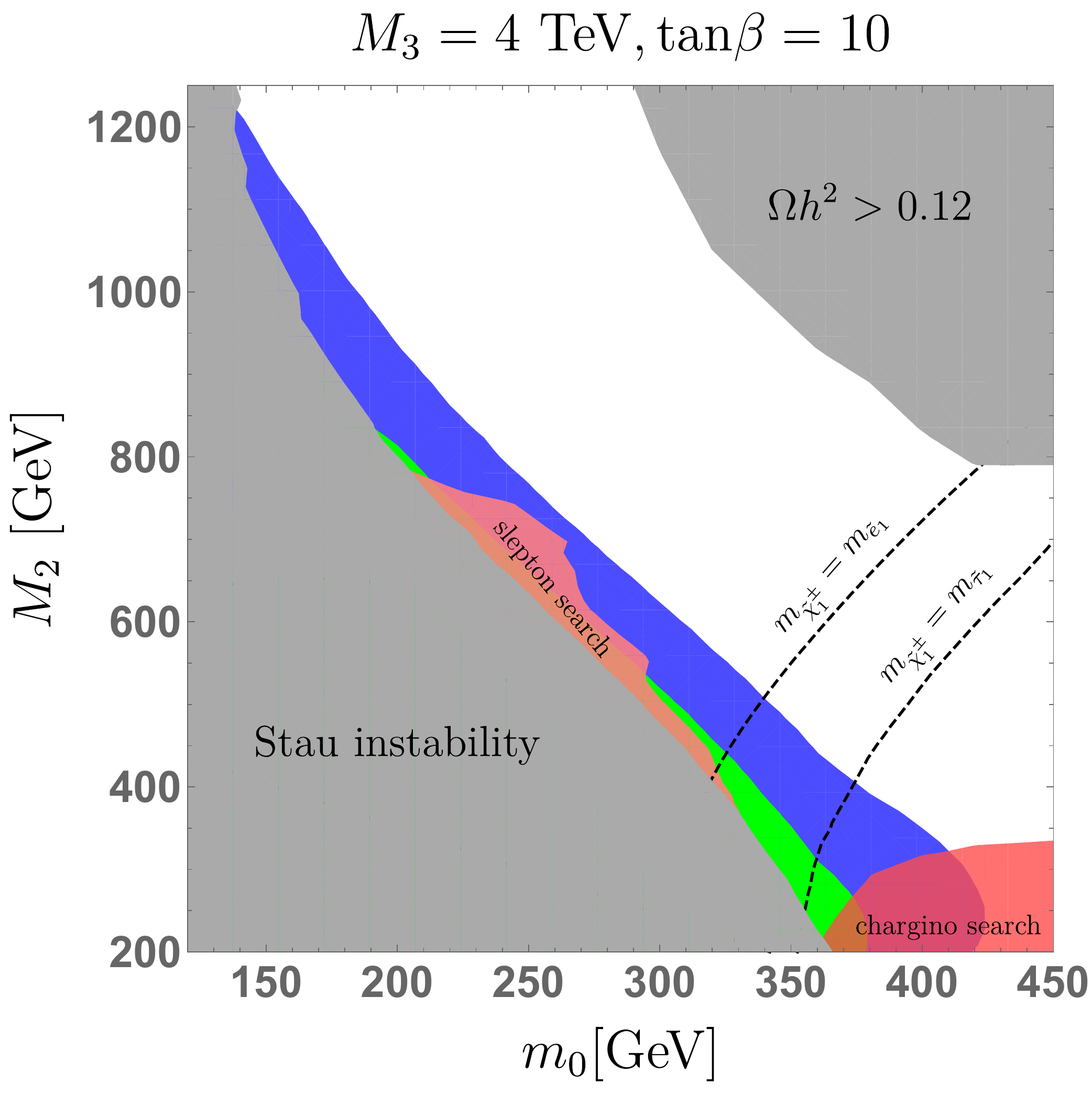}
  \caption{Results with universal soft scalar masses. The dark matter relic density is satisfied everywhere except the grey regions (see text for details). The green (blue) regions are consistent with the muon $g-2$ at $1(2)\sigma$. The pink and red regions are excluded by ATLAS slepton~\cite{1803.02762} and chargino~\cite{1712.08119} searches respectively.}
  \label{fig:m0-results}
\end{figure}

The results of this scan are shown in fig.~\ref{fig:m0-results}, where we have fixed $M_3=4\,$TeV in order to obtain approximately the correct Higgs mass, and taken $\tan\beta=10$ and $\mu>0$.
$M_1$ has been varied throughout the parameter space in order to obtain the correct dark matter relic density. 
In the grey regions there is no value of $M_1$ that yields a consistent spectrum and the correct relic density with a bino-like LSP.
At small $m_0$, this is due to the fact that the relatively large value of $\mu$ ($\sim4\,$TeV) causes the lightest stau to become tachyonic; 
the strongest constraint here comes from vacuum stability due to the presence of charge-breaking minima in the scalar potential~\cite{hep-ph/9507294, hep-ph/9612464, 1011.0260}.
It is this constraint that prevents the minimal model with $m_0=0$ from being able to explain both the dark matter abundance with a bino LSP and the muon $g-2$. 
In the upper-right grey region, the stau becomes too heavy for $\tilde{B}-\tilde{\tau}$ co-annihilation to sufficiently reduce the relic density of the bino dark matter. 
In the remaining parameter space the correct relic abundance is obtained via bino-stau and/or bino-wino co-annihilations, which require an $\mathcal{O}(10)\,$GeV mass-splitting between the bino and its co-annihilation partner~\cite{hep-ph/0206266, 1311.2162, 1403.0715}. 

The green (blue) regions in fig.~\ref{fig:m0-results} are also consistent with the observed value of the muon $g-2$ at $1(2)\sigma$. 
There are then viable regions of parameter space that can account both for the dark matter abundance and the muon $g-2$; however, these regions also feature relatively light selectrons/smuons that can be produced at the LHC. 
The pink region is already excluded by slepton searches with $36\,\text{fb}^{-1}$ integrated luminosity~\cite{1803.02762, 1806.05264}. 
There are also dedicated searches by ATLAS and CMS for the compressed spectra expected in co-annihilation scenarios, and the red region is excluded by compressed chargino searches~\cite{1712.08119, CMS-PAS-SUS-16-025}. 
This leaves a small region of viable parameter space that can explain the muon $g-2$ at $1\sigma$. 
Note this is also where both $\tilde{B}-\tilde{\tau}$ and $\tilde{B}-\tilde{W}$ co-annihilations are active. 
This surviving region escapes existing LHC searches due to the opening of additional decay modes: $\chi_1^\pm$ and $\chi_2^0$ decay dominantly to tau final states, while the left-handed sleptons decay into neutrinos ($\tilde{l}^\pm\to\chi_1^\pm\nu$) with a branching ratio of $\sim60\%$.
Nevertheless, this region should also be tested by slepton searches in the future. 
A benchmark point that satisfies all of the current constraints is shown in tab.~\ref{tab:m0-benchmark}; note that the coloured sparticles are all decoupled and well beyond the reach of the LHC. 

Finally, let us comment on the effect of $\tan\beta$, which has been fixed in fig.~\ref{fig:m0-results}. 
It should be clear from eq.~\eqref{eq:g-2} that increasing $\tan\beta$ moves the best-fit region for the $g-2$ to heavier smuon masses, and hence larger values of $m_0$. 
However, increasing $\tan\beta$ also leads to a stronger bound from stau instability, and this eventually excludes all of the $2\sigma$ best-fit region for $\tan\beta>20$. 
For small values of $\tan\beta$ it becomes difficult to obtain the observed Higgs mass. 

\begin{table}[t]
\centering
\begin{tabu}{|c|c|c|}
  \hline
  $M_1(M_\text{GUT})$ & 396 \\
  $M_2(M_\text{GUT})$ & 310 \\
  $M_3(M_\text{GUT})$ & 4000 \\
  $m_0(M_\text{GUT})$ & 350 \\
  $\tan\beta$ & 10 \\
  \hline
  $m_{\tilde{g}}$ & 7940 \\
  $m_{\tilde{q}}$ & 6700 \\
  $m_{\tilde{t}_1}$, $m_{\tilde{t}_2}$ & 5840, 6300 \\
  $m_{\tilde{l}_L}$, $m_{\tilde{l}_R}$ & 267, 368 \\
  $m_{\tilde{\tau}_1}$, $m_{\tilde{\tau}_2}$ & 166, 421 \\
  $m_{\tilde{\nu}}$ & 255 \\
  $m_{\tilde{\chi}^0_1}$ & 148 \\
  $m_{\tilde{\chi}^\pm_1} \simeq m_{\tilde{\chi}^0_2}$ & 189 \\
  $m_{\tilde{\chi}^\pm_2} \simeq m_{\tilde{\chi}^0_{3,4}}$ & 4150 \\
  $m_A \simeq m_{H^0} \simeq M_{H^\pm}$ & 4150 \\
  \hline
  $m_h$ & 123.3 \\
  $\Delta a_\mu$ &  2.27$\times10^{-9}$ \\
  $\Omega_{DM}h^2$ & 0.118 \\
  \hline
\end{tabu}
\caption{Complete spectrum for a benchmark point that satisfies all existing constraints. All masses are in GeV.}
\label{tab:m0-benchmark}
\end{table}

%%%%%%%%%%%%%%%%%%%%%%%%%%%%%%%%%%

\section{Gaugino + Higgs mediation} \label{sec:Higgs-med}

It should be clear from the previous section that a significant challenge to explaining both the muon $g-2$ and dark matter in gaugino mediation models is the presence of the stau instability. 
A potential solution to this problem would be to split the slepton spectrum and give large soft masses to the third generation; however, such non-universal slepton masses generically give rise to lepton flavour violation, with potentially large effects in $\mu\to e\gamma$ in the regions of parameter space relevant for explaining the muon $g-2$~\cite{1309.3065}. 
What is needed, therefore, is a mechanism to split the third generation sfermions without inducing large FCNC; this can be provided by Higgs mediation.

\subsection{Model}

\begin{figure}
  \centering
  \includegraphics[width=0.45\textwidth]{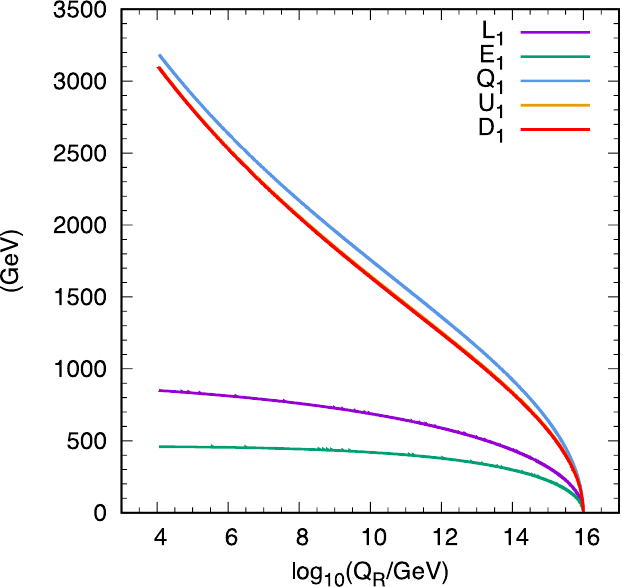}
  \includegraphics[width=0.45\textwidth]{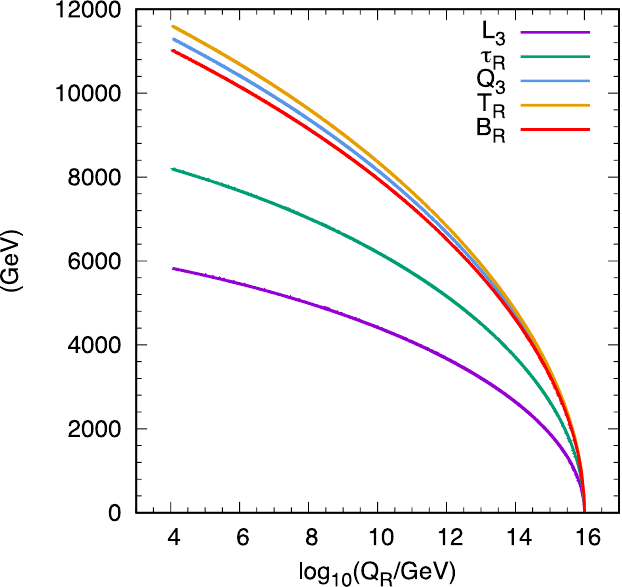}
  \caption{RG evolution of the sfermion soft masses in gaugino+Higgs mediation. The left (right) panels are for the first (third) generation (the smuon masses are only slightly heavier than the selectron masses). We have taken $M_1=440$\,GeV, $M_2=220$\,GeV, $M_3=-2000$\,GeV, $m_{H_{u/d}}^2=-7\times10^8\,\text{(GeV)}^2$, and $\tan\beta=40$.}
  \label{fig:RGE}
\end{figure}

In gaugino+Higgs mediation, only the Higgs doublets and gauginos couple to the SUSY breaking field. 
The soft SUSY breaking squared masses for the Higgs doublets are assumed to be negative and quite large, $\mathcal{O}(10^8$-$10^9)$\,GeV$^2$.  
Then, Higgs loops generate large positive squared masses for the third generation sfermions at the one-loop level~\cite{Yin:2016shg,Yanagida:2016kag,Yamaguchi:2016oqz}. 
For the first and second generation these contributions are suppressed by their small Yukawa couplings. 
Consequently, the third generation sfermions naturally become much heavier than the first/second generation without inducing too large FCNC~\cite{Yanagida:2018eho}. The absence of large FCNC is understood from the fact that gaugino+Higgs mediation belongs to the framework of minimal flavor violation~\cite{hep-ph/0007085}. 
The first/second generation sfermions do receive masses from the gauginos at one-loop, and non-Yukawa suppressed contributions from the Higgs at two-loops. These positive contributions from Higgs loops to the slepton squared masses are particularly important~\cite{Yin:2016shg,Yanagida:2016kag}, allowing the sleptons to be heavier than the bino (and the wino when its mass is small).
This RG running of the soft masses for the first and third generation sfermions is exemplified in fig.~\ref{fig:RGE}, with $Q_R$ being a renormalisation scale.  
The resulting split spectrum is advantageous both to explain the muon $g-2$ while evading the constraint from vacuum stability in the stau-Higgs potential, and to raise the Higgs mass to be consistent with its measured value.
Note that $|\mu|^2+m_{H_{u,d}}^2$ remain positive during the renormalisation group (RG) running for large $\tan\beta$, and there are no tachyonic directions (including the $D$-flat direction) in the Higgs potential at energies higher than the stop mass scale. 

The K\"ahler potential once again takes the form
\begin{equation}
  K = -3 M_P^2 \ln \left[1 - \frac{f(Z, Z^\dag) + Q_I^\dag Q_I + \Delta K'}{3M_P^2}\right], \, \\
\end{equation}
with $\Delta K'$ now given by
\begin{equation}
  \Delta K' = d' \frac{|Z|^2}{M_P^2} (H_u^\dag H_u + \kappa H_d^\dag H_d) - c_b \frac{|Z|^2}{M_P^2} H_u H_d + h.c. \,,
\end{equation}
and where $d'$ is assumed to be positive and $\kappa=1$. 
Then, the Higgs doublets obtain soft masses
\begin{equation}
  m_{H_u}^2 = m_{H_d}^2 = - d' \frac{|F_Z|^2}{M_P^2} = - 3 d' \left(\frac{\partial^2 K}{\partial Z \partial Z^\dag}\right)^{-1} m_{3/2}^2 \,.
\end{equation}
The Higgs B-term is the same as in eq.~\eqref{eq:higgs_b-term}. 
Note that the sfermion masses vanish at tree-level with the above K\"ahler potential, which is a key ingredient to avoid too large FCNC. 

There will also be contributions to the soft masses coming from anomaly mediation; however these are flavour-blind and do not affect the above conclusion. 
These contributions are also numerically rather small; for example, anomaly mediation gives slepton and wino masses $\sqrt{|m_{\tilde{e}_L}^2|} \sim M_2 \sim g_2^2/(16 \pi^2) \,m_{3/2} \approx 80\,{\rm GeV} (m_{3/2}/30\,{\rm TeV})$. 
We neglect these contributions in the following, as they are not expected to significantly affect our results.

%%%%%%%%%%%%%%%%%%%%%%%%%%%%%%%%%%

\subsection{Results}

For gaugino+Higgs mediation we then have the following free parameters:
\begin{equation}
  M_1, M_2, M_3, m_{H_u}^2, \tan\beta, \text{sgn}(\mu) \,,
\end{equation}
where we have assumed $m_{H_u}^2=m_{H_d}^2$ at the GUT scale. 
Achieving viable EWSB in this setup requires $\mu$ to be large ($\mathcal{O}(10)\,$TeV) and $m_{H_d}^2-m_{H_u}^2\gtrsim0$ at low scales~\cite{Yin:2016shg}. 
This second condition forces one towards large values of $\tan\beta$ in order to enhance the down-type Yukawa couplings appearing in the $m_{H_d}^2$ beta function, such that $\beta_{m_{H_u}^2} > \beta_{m_{H_d}^2}$ (note that $\beta_{m_{H_u}^2}$ and $\beta_{m_{H_d}^2}$ are both negative). 
At large $\tan\beta$ the bottom Yukawa coupling also receives significant threshold corrections~\cite{hep-ph/9912516} (analogous to those for $y_\mu$ relevant for the $g-2$ in eq.~\eqref{eq:Delta}). 
If $\mu M_3<0$ these corrections increase $y_b$, allowing one to have viable EWSB with somewhat smaller values of $\tan\beta$; we will take $M_3<0$ and $\mu>0$ in the following. 

\begin{figure}[t]
  \centering
  \includegraphics[width=0.6\textwidth]{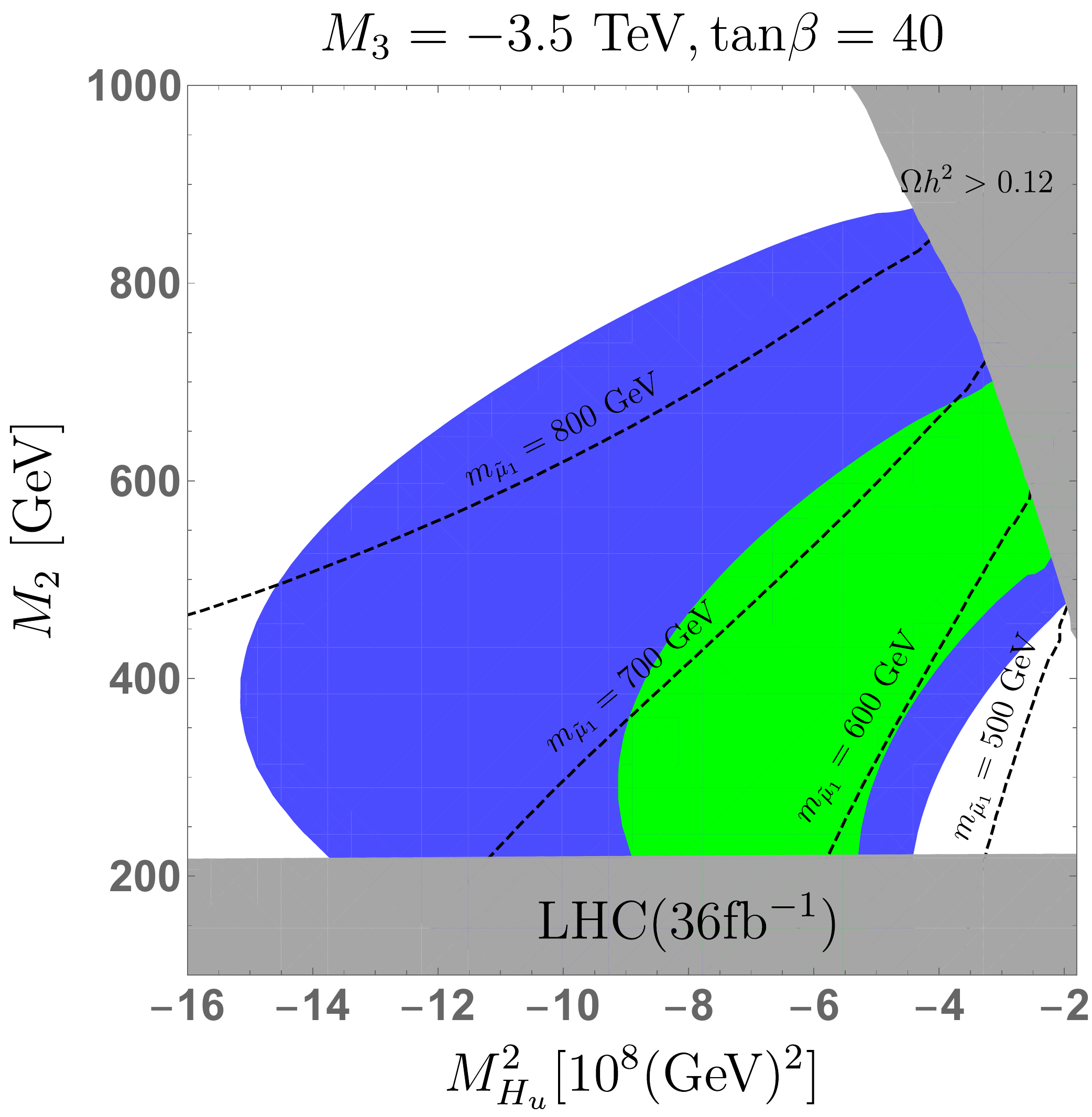}
  \caption{Region of parameter space where the dark matter abundance is obtained via $\tilde{B}-\tilde{W}$ co-annihilation. The green (blue) regions are consistent with the muon $g-2$ at $1(2)\sigma$. The lower grey region is excluded by compressed chargino searches~\cite{1712.08119}. In the upper grey region it is not possible to obtain the correct relic density.}
  \label{fig:bino-wino}
\end{figure}

Within this gaugino+Higgs mediation setup, one can show that it is possible to obtain spectra suitable for producing the dark matter abundance via either bino-wino or bino-slepton co-annihilations. 
Let us begin by focusing on the former case; the relevant region of parameter space is shown in fig.~\ref{fig:bino-wino}, where we have taken $\tan\beta=40$ and $M_3=-3.5\,$TeV. 
$M_1$ has been varied in order to achieve a small mass splitting between the bino-like LSP and wino-like next-to-lightest supersymmetric particle (NLSP) and obtain the correct relic abundance through co-annihilations. 
First, notice that dark matter masses below $\sim200$\,GeV are already excluded by compressed chargino searches~\cite{1712.08119, CMS-PAS-SUS-16-025}, while in the upper right corner of parameter space it is no longer possible to obtain the observed dark matter abundance with a bino LSP. 
Nevertheless, there remain significant regions of parameter space that can account for the observed dark matter abundance and are consistent with the muon $g-2$. 
The required smuon masses are around $600-700$\,GeV in the $1\sigma$ best-fit region for the $g-2$; these are safe from current slepton searches, however it may be possible to test at least some of this region at the HL-LHC. 
A benchmark point is given in tab.~\ref{tab:BPs}.

\begin{figure}[t]
  \centering
  \includegraphics[width=0.6\textwidth]{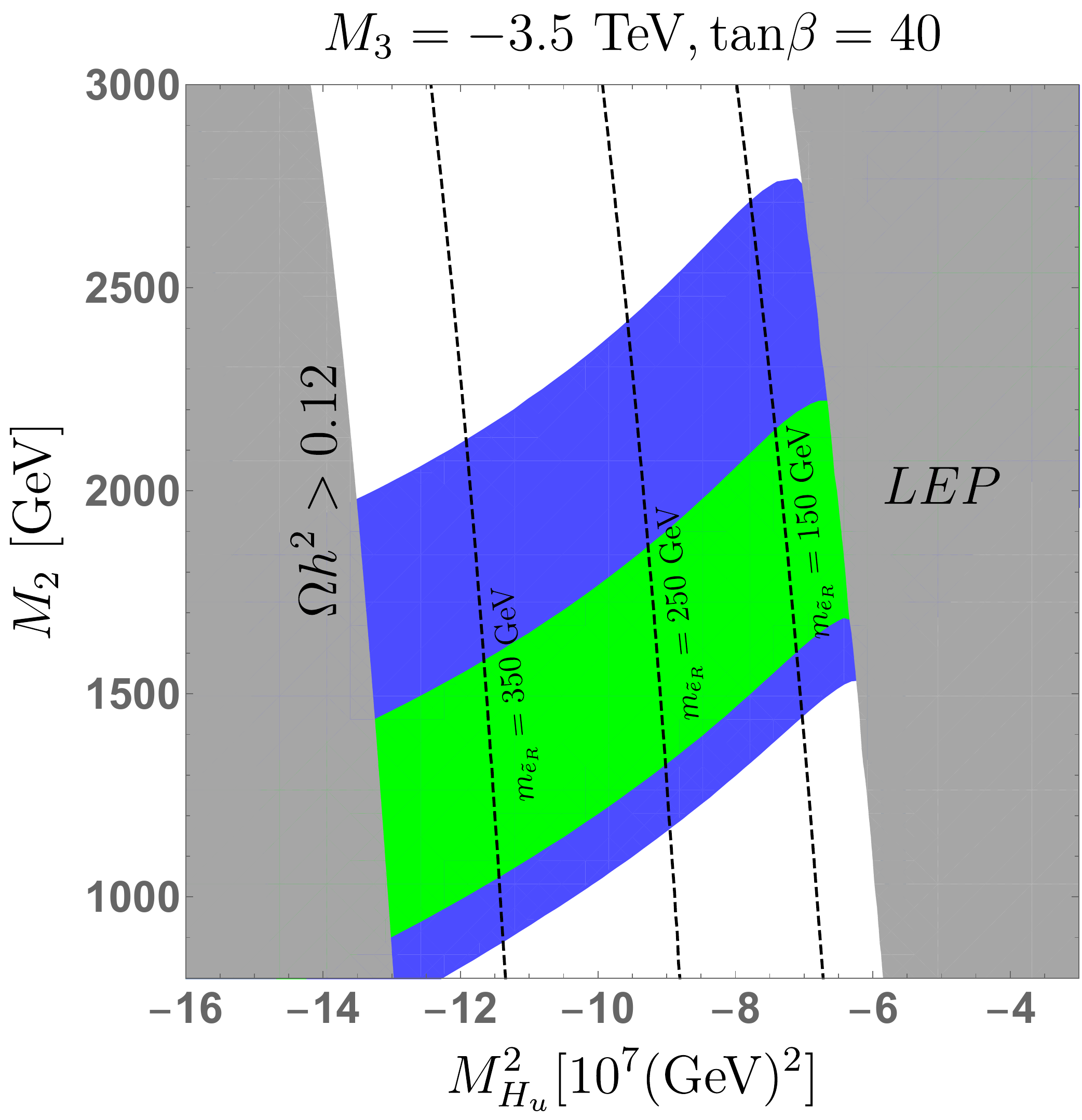}
  \caption{Region of parameter space where the dark matter abundance is obtained via $\tilde{B}-\tilde{l}$ co-annihilation. The green (blue) regions are consistent with the muon $g-2$ at $1(2)\sigma$. In the left grey region the sleptons become too heavy to achieve the correct relic abundance; the right grey region is excluded by slepton searches at LEP~\cite{LEP-slepton}.}
  \label{fig:bino-slepton}
\end{figure}

Let us turn now to the case of bino-slepton co-annihilation, which can be realised in the region of parameter space shown in fig.~\ref{fig:bino-slepton}. 
Recall that the third generation sleptons are decoupled in this framework, and hence the lightest slepton and co-annihilation partner is the right-handed selectron. 
There is a lower limit on the dark matter mass due to slepton searches at LEP, which for a 10\,GeV mass splitting between $\tilde{e}_R$ and the LSP give the constraint $m_{\tilde{e}_R}\gtrsim96\,$GeV~\cite{LEP-slepton}. 
There is also an upper bound on the dark matter mass of around 400\,GeV, beyond which co-annihilations are no longer sufficiently effective in reducing the relic abundance. 
Fortunately, it turns out that the muon $g-2$ can also be easily explained within this mass range. 
Despite the fact that this scenario features relatively light sleptons, these are extremely difficult to probe at the LHC due to both the compressed spectrum and the fact that the left-handed sleptons, which have a much larger production cross-section, are significantly heavier. 
However, the right-handed sleptons could potentially be discovered at a future International Linear Collider. 
Furthermore, within the best fit region for the muon $g-2$, the mass of the lightest chargino is always below around 2\,TeV. 
While LHC electroweakino searches are not currently sensitive to the best-fit region, these may be the best way to probe this scenario at the LHC in the near future. 
A benchmark point is again provided in tab.~\ref{tab:BPs}.
Finally, note that $\tan\beta$ is predicted to be around 40 in gaugino+Higgs mediation for $\mu M_3<0$; larger $\tan\beta$ leads to a Landau pole below the GUT scale due to too large $y_b$, and smaller $\tan\beta$ leads to unsuccessful EWSB (CP-odd Higgs becomes tachyonic). 
The viable region of $\tan\beta$ depends slightly on $M_3$ through the radiative correction to $y_b$.

\begin{table}[t]
\centering
\begin{tabu}{|c|c|c|}
  \hline
  & BP1 ($\tilde{B}-\tilde{W}$) & BP2 ($\tilde{B}-\tilde{l}$) \\
  \hline
  $M_1(M_\text{GUT})$ & 810.4 & 473.5 \\
  $M_2(M_\text{GUT})$ & 400 & 1500 \\
  $M_3(M_\text{GUT})$ & -3500 & -3500 \\
  $m_{H_u}^2=m_{H_d}^2(M_\text{GUT})$ & $-6.0\times10^8\,\text{GeV}^2$ & $-9.0\times10^7\,\text{GeV}^2$ \\
  $\tan\beta$ & 40 & 40 \\
  \hline
  $m_{\tilde{g}}$ & 7230 & 7100 \\
  $m_{\tilde{q}}$ & 5800 & 6000 \\
  $m_{\tilde{t}_1}$, $m_{\tilde{t}_2}$ & 11600, 11700 & 6680, 6770 \\
  $m_{\tilde{e}_L}$, $m_{\tilde{e}_R}$ & 704, 653 & 961, 246 \\
  $m_{\tilde{\mu}_L}$, $m_{\tilde{\mu}_R}$ & 742, 733 & 968, 296 \\
  $m_{\tilde{\tau}_1}$, $m_{\tilde{\tau}_2}$ & 5500, 7780 & 2410, 3200 \\
  $m_{\tilde{\nu}_e}$, $m_{\tilde{\nu}_\mu}$, $m_{\tilde{\nu}_\tau}$ & 700, 738, 5500 & 958, 965, 2430 \\
  $m_{\tilde{\chi}^0_1}$ & 395 & 241 \\
  $m_{\tilde{\chi}^\pm_1} \simeq m_{\tilde{\chi}^0_2}$ & 422 & 1360 \\
  $m_{\tilde{\chi}^\pm_2} \simeq m_{\tilde{\chi}^0_{3,4}}$ & 20800 & 8540 \\
  $m_A \simeq m_{H^0} \simeq M_{H^\pm}$ & 6560 & 2170 \\
  \hline
  $m_h$ & 124.9 & 124.8 \\
  $\Delta a_\mu$ & $2.64\times10^{-9}$ & $2.87\times10^{-9}$ \\
  $\Omega_{DM}h^2$ & 0.1195 & 0.1188 \\
  \hline
\end{tabu}
\caption{Complete spectra for two benchmark points corresponding to $\tilde{B}-\tilde{W}$ (BP1) and $\tilde{B}-\tilde{l}$ (BP2) co-annihilation. All masses are in GeV.}
\label{tab:BPs}
\end{table}

%%%%%%%%%%%%%%%%%%%%%%%%%%%%%%%%%%

\section{Conclusion}

The observed discrepancy in the muon $g-2$ remains an intriguing hint for new physics that can be naturally explained in the context of supersymmetry. 
Furthermore, the required electroweak superpartners can also play a role in producing the dark matter abundance, in particular via co-annihilations with a bino-like LSP.
In this work, we have shown how this scenario can be realised within a realistic UV setup based on gaugino mediation. 

However, we find that the constraint from stau instability immediately forces one to go beyond the minimal model with only non-universal gaugino masses at the GUT scale. 
We therefore focused on two simple extensions of this minimal setup. 
Firstly, we allowed for an additional universal soft scalar mass (a similar effect could arise from $B-L$ gaugino mediation). 
This model can simultaneously explain both the muon $g-2$ and the dark matter abundance; however, stau vacuum stability still provides a significant restriction on the parameter space. 
Furthermore, the remaining best-fit regions are already being tested by direct searches for the light sleptons or compressed chargino/neutralino. 

We then considered a model with gaugino and Higgs mediation, where tachyonic soft masses for the Higgs doublets lead to a splitting of the third generation sfermions from the first two generations via renormalisation group running. 
This allows one to easily evade the constraint from stau instability without inducing too large FCNC. 
We showed that within this framework there are viable regions of parameter space that can explain the muon $g-2$, with the dark matter abundance obtained via either bino-wino or bino-slepton co-annihilation. 
In the former case, compressed chargino searches currently require $m_{\chi^0_1}\gtrsim200$\,GeV, and LHC slepton searches may also be sensitive to some of the best-fit region in the future. 
In the latter case, the right-handed selectron is the NLSP and co-annihilation partner; this leads to an upper bound on the dark matter mass of 400\,GeV, beyond which co-annihilations cannot sufficiently reduce the relic abundance. 
The light right-handed sleptons are inaccessible at the LHC due to their small cross-section and compressed spectrum; however the wino may also remain light enough to be probed by direct searches in the future. 
In both cases, a future International Linear Collider could also test some of the viable regions via production of the co-annihilation partners. 

%%%%%%%%%%%%%%%%%%%%%%%%%%%%%%%%%%

\section*{Acknowledgements}
This work is supported by Grants-in-Aid for Scientific Research from the Ministry of Education, Culture, Sports, Science, and Technology (MEXT), Japan, No. 26104001 (T.T.Y.), No. 26104009 (T.T.Y.), No. 16H02176 (T.T.Y.), No. 17H02878 (T.T.Y.), No. 15H05889 (N.Y.),
No. 15K21733 (N.Y.), No. 17H05396 (N.Y.) and No. 17H02875 (N.Y.), and by the World Premier International Research Center Initiative (WPI), MEXT, Japan (P.C., C.H. and T.T.Y.). 

%%%%%%%%%%%%%%%%%%%%%%%%%%%%%%%%%%

\appendix
\section{$B-L$ gaugino mediation} \label{app:B-L}

In this appendix, we show how the non-zero sfermion masses arise in a $B-L$ extension of the MSSM. 
The superpotential is given by
\begin{align}
  W &= y_e H_d L \bar E + y_u Q H_u \bar U + y_d H_d Q \bar D + y_\nu L H_u \bar N + \frac{1}{2} \lambda_S S \bar N \bar N \nonumber \\
  &+ \lambda_X X (S \bar S - v_{B-L}^2) \,,
\end{align}
where we omit flavor indices; $\bar N$ is a right-handed neutrino superfield. Here, $S$, $\bar S$ and $X$ are singlets under the MSSM gauge group.
The $U(1)_{B-L}$ charges are summarised in tab. \ref{tab:charges}.
The soft SUSY breaking masses are generated via the RG equation,
\begin{equation}
  \frac{d m_{Q_I}^2}{d \ln Q_R} \ni \frac{g_{B-L}^2}{16\pi^2} ( - 8 q_{B-L,I}^2  M_{B-L}^2 ) \,,
\end{equation}
where $g_{B-L}$ is the $U(1)_{B-L}$ gauge coupling, $q_{B-L,I}$ is the $U(1)_{B-L}$ charge of an MSSM superfield $Q_I$, and $M_{B-L}$ is the mass of the $B-L$ gaugino. 
We assume kinetic mixing between $U(1)_Y$ and $U(1)_{B-L}$ is negligible. 
Integrating the RG equation, we obtain
\begin{equation}
  m_{Q_I}^2 (M_{\rm break}) \simeq \frac{g_{B-L}^2}{2\pi^2} q_{B-L,I}^2  M_{B-L}^2 \ln \frac{M_{\rm GUT}}{M_{\rm break}} \,,
\end{equation}
where $M_{\rm break}$ is the $U(1)_{B-L}$ breaking scale of $\sim g_{B-L} v_{B-L}$, which is assumed to be close to the GUT scale $M_{\rm GUT}$.

\begin{table}[t]
  \centering
  \begin{tabu}{|c|c|c|c|c|c|c|c|c|c|c|}
    \hline
     $Q$ & $\bar U$ & $\bar D$ & $L$ & $\bar E$ & $\bar N$ &$H_u$ & $H_d$ & $S$ & $\bar S$ & $X$\\
    \hline
    $1/3$ & $-1/3$ & $-1/3$ & $-1$ & $1$ & $1$ &$0$&$0$& $-2$ & $2$ & 0 \\
    \hline
  \end{tabu}
  \caption{Charge assignment of $U(1)_{B-L}$.}
  \label{tab:charges}
\end{table}

%%%%%%%%%%%%%%%%%%%%%%%%%%%%%%%%%%

\bibliographystyle{jhep}
\bibliography{draft}

\providecommand{\href}[2]{#2}\begingroup\raggedright\begin{thebibliography}{10}

\bibitem{hep-ex/0602035}
{\bf Muon g-2} Collaboration, G.~W. Bennett et~al., {\it {Final Report of the
  Muon E821 Anomalous Magnetic Moment Measurement at BNL}},  {\em Phys. Rev.}
  {\bf D73} (2006) 072003, [\href{http://arxiv.org/abs/hep-ex/0602035}{{\tt
  hep-ex/0602035}}].

\bibitem{1701.02807}
{\bf Muon g-2} Collaboration, A.~Chapelain, {\it {The Muon g-2 experiment at
  Fermilab}},  {\em EPJ Web Conf.} {\bf 137} (2017) 08001,
  [\href{http://arxiv.org/abs/1701.02807}{{\tt arXiv:1701.02807}}].

\bibitem{1805.01607}
G.~Bhattacharyya, T.~T. Yanagida, and N.~Yokozaki, {\it {An extended gauge
  mediation for muon $(g-2)$ explanation}},  {\em Phys. Lett.} {\bf B784}
  (2018) 118--121, [\href{http://arxiv.org/abs/1805.01607}{{\tt
  arXiv:1805.01607}}].

\bibitem{1404.4841}
M.~Chakraborti, U.~Chattopadhyay, A.~Choudhury, A.~Datta, and S.~Poddar, {\it
  {The Electroweak Sector of the pMSSM in the Light of LHC - 8 TeV and Other
  Data}},  {\em JHEP} {\bf 07} (2014) 019,
  [\href{http://arxiv.org/abs/1404.4841}{{\tt arXiv:1404.4841}}].

\bibitem{1406.6925}
S.~P. Das, M.~Guchait, and D.~P. Roy, {\it {Testing SUSY models for the muon
  g-2 anomaly via chargino-neutralino pair production at the LHC}},  {\em Phys.
  Rev.} {\bf D90} (2014), no.~5 055011,
  [\href{http://arxiv.org/abs/1406.6925}{{\tt arXiv:1406.6925}}].

\bibitem{1409.3930}
T.~Li and S.~Raza, {\it {Electroweak supersymmetry from the generalized minimal
  supergravity model in the MSSM}},  {\em Phys. Rev.} {\bf D91} (2015), no.~5
  055016, [\href{http://arxiv.org/abs/1409.3930}{{\tt arXiv:1409.3930}}].

\bibitem{1503.08219}
K.~Kowalska, L.~Roszkowski, E.~M. Sessolo, and A.~J. Williams, {\it
  {GUT-inspired SUSY and the muon $g-2$ anomaly: prospects for LHC 14 TeV}},
  {\em JHEP} {\bf 06} (2015) 020, [\href{http://arxiv.org/abs/1503.08219}{{\tt
  arXiv:1503.08219}}].

\bibitem{1504.00505}
F.~Wang, W.~Wang, and J.~M. Yang, {\it {Reconcile muon g-2 anomaly with LHC
  data in SUGRA with generalized gravity mediation}},  {\em JHEP} {\bf 06}
  (2015) 079, [\href{http://arxiv.org/abs/1504.00505}{{\tt arXiv:1504.00505}}].

\bibitem{1505.05877}
B.~P. Padley, K.~Sinha, and K.~Wang, {\it {Natural Supersymmetry, Muon $g-2$,
  and the Last Crevices for the Top Squark}},  {\em Phys. Rev.} {\bf D92}
  (2015), no.~5 055025, [\href{http://arxiv.org/abs/1505.05877}{{\tt
  arXiv:1505.05877}}].

\bibitem{1505.05896}
M.~A. Ajaib, B.~Dutta, T.~Ghosh, I.~Gogoladze, and Q.~Shafi, {\it {Neutralinos
  and sleptons at the LHC in light of muon $(g-2)_{\mu}$}},  {\em Phys. Rev.}
  {\bf D92} (2015), no.~7 075033, [\href{http://arxiv.org/abs/1505.05896}{{\tt
  arXiv:1505.05896}}].

\bibitem{1507.01395}
M.~Chakraborti, U.~Chattopadhyay, A.~Choudhury, A.~Datta, and S.~Poddar, {\it
  {Reduced LHC constraints for higgsino-like heavier electroweakinos}},  {\em
  JHEP} {\bf 11} (2015) 050, [\href{http://arxiv.org/abs/1507.01395}{{\tt
  arXiv:1507.01395}}].

\bibitem{1608.03641}
A.~Kobakhidze, M.~Talia, and L.~Wu, {\it {Probing the MSSM explanation of the
  muon g-2 anomaly in dark matter experiments and at a 100 TeV $pp$ collider}},
   {\em Phys. Rev.} {\bf D95} (2017), no.~5 055023,
  [\href{http://arxiv.org/abs/1608.03641}{{\tt arXiv:1608.03641}}].

\bibitem{1704.05287}
M.~Endo, K.~Hamaguchi, S.~Iwamoto, and K.~Yanagi, {\it {Probing minimal SUSY
  scenarios in the light of muon $g-2$ and dark matter}},  {\em JHEP} {\bf 06}
  (2017) 031, [\href{http://arxiv.org/abs/1704.05287}{{\tt arXiv:1704.05287}}].

\bibitem{1710.11091}
E.~Bagnaschi et~al., {\it {Likelihood Analysis of the pMSSM11 in Light of LHC
  13-TeV Data}},  {\em Eur. Phys. J.} {\bf C78} (2018), no.~3 256,
  [\href{http://arxiv.org/abs/1710.11091}{{\tt arXiv:1710.11091}}].

\bibitem{1211.4873}
C.~Cheung, L.~J. Hall, D.~Pinner, and J.~T. Ruderman, {\it {Prospects and Blind
  Spots for Neutralino Dark Matter}},  {\em JHEP} {\bf 05} (2013) 100,
  [\href{http://arxiv.org/abs/1211.4873}{{\tt arXiv:1211.4873}}].

\bibitem{1404.0392}
P.~Huang and C.~E.~M. Wagner, {\it {Blind Spots for neutralino Dark Matter in
  the MSSM with an intermediate $m_A$}},  {\em Phys. Rev.} {\bf D90} (2014),
  no.~1 015018, [\href{http://arxiv.org/abs/1404.0392}{{\tt arXiv:1404.0392}}].

\bibitem{1612.02387}
T.~Han, F.~Kling, S.~Su, and Y.~Wu, {\it {Unblinding the dark matter blind
  spots}},  {\em JHEP} {\bf 02} (2017) 057,
  [\href{http://arxiv.org/abs/1612.02387}{{\tt arXiv:1612.02387}}].

\bibitem{1701.02737}
P.~Huang, R.~A. Roglans, D.~D. Spiegel, Y.~Sun, and C.~E.~M. Wagner, {\it
  {Constraints on Supersymmetric Dark Matter for Heavy Scalar Superpartners}},
  {\em Phys. Rev.} {\bf D95} (2017), no.~9 095021,
  [\href{http://arxiv.org/abs/1701.02737}{{\tt arXiv:1701.02737}}].

\bibitem{1303.3040}
T.~Han, Z.~Liu, and A.~Natarajan, {\it {Dark matter and Higgs bosons in the
  MSSM}},  {\em JHEP} {\bf 11} (2013) 008,
  [\href{http://arxiv.org/abs/1303.3040}{{\tt arXiv:1303.3040}}].

\bibitem{hep-ph/0106275}
G.~Belanger, F.~Boudjema, A.~Cottrant, R.~M. Godbole, and A.~Semenov, {\it {The
  MSSM invisible Higgs in the light of dark matter and g-2}},  {\em Phys.
  Lett.} {\bf B519} (2001) 93--102,
  [\href{http://arxiv.org/abs/hep-ph/0106275}{{\tt hep-ph/0106275}}].

\bibitem{hep-ph/0511034}
H.~Baer, T.~Krupovnickas, A.~Mustafayev, E.-K. Park, S.~Profumo, and X.~Tata,
  {\it {Exploring the BWCA (bino-wino co-annihilation) scenario for neutralino
  dark matter}},  {\em JHEP} {\bf 12} (2005) 011,
  [\href{http://arxiv.org/abs/hep-ph/0511034}{{\tt hep-ph/0511034}}].

\bibitem{hep-ph/9905481}
J.~R. Ellis, T.~Falk, K.~A. Olive, and M.~Srednicki, {\it {Calculations of
  neutralino-stau coannihilation channels and the cosmologically relevant
  region of MSSM parameter space}},  {\em Astropart. Phys.} {\bf 13} (2000)
  181--213, [\href{http://arxiv.org/abs/hep-ph/9905481}{{\tt hep-ph/9905481}}].
  [Erratum: Astropart. Phys.15,413(2001)].

\bibitem{1805.02802}
P.~Cox, C.~Han, and T.~T. Yanagida, {\it {Muon $g-2$ and dark matter in the
  minimal supersymmetric standard model}},  {\em Phys. Rev.} {\bf D98} (2018),
  no.~5 055015, [\href{http://arxiv.org/abs/1805.02802}{{\tt
  arXiv:1805.02802}}].

\bibitem{Inoue:1991rk}
K.~Inoue, M.~Kawasaki, M.~Yamaguchi, and T.~Yanagida, {\it {Vanishing squark
  and slepton masses in a class of supergravity models}},  {\em Phys. Rev.}
  {\bf D45} (1992) 328--337.

\bibitem{hep-ph/9911293}
D.~E. Kaplan, G.~D. Kribs, and M.~Schmaltz, {\it {Supersymmetry breaking
  through transparent extra dimensions}},  {\em Phys. Rev.} {\bf D62} (2000)
  035010, [\href{http://arxiv.org/abs/hep-ph/9911293}{{\tt hep-ph/9911293}}].

\bibitem{hep-ph/9911323}
Z.~Chacko, M.~A. Luty, A.~E. Nelson, and E.~Ponton, {\it {Gaugino mediated
  supersymmetry breaking}},  {\em JHEP} {\bf 01} (2000) 003,
  [\href{http://arxiv.org/abs/hep-ph/9911323}{{\tt hep-ph/9911323}}].

\bibitem{1501.07447}
K.~Harigaya, T.~T. Yanagida, and N.~Yokozaki, {\it {Higgs boson mass of
  125 GeV and $g-2$ of the muon in a gaugino mediation model}},  {\em Phys.
  Rev.} {\bf D91} (2015), no.~7 075010,
  [\href{http://arxiv.org/abs/1501.07447}{{\tt arXiv:1501.07447}}].

\bibitem{Yin:2016shg}
W.~Yin and N.~Yokozaki, {\it {Splitting mass spectra and muon $g-2$ in
  Higgs-anomaly mediation}},  {\em Phys. Lett.} {\bf B762} (2016) 72--79,
  [\href{http://arxiv.org/abs/1607.05705}{{\tt arXiv:1607.05705}}].

\bibitem{Yanagida:2016kag}
T.~T. Yanagida, W.~Yin, and N.~Yokozaki, {\it {Nambu-Goldstone Boson Hypothesis
  for Squarks and Sleptons in Pure Gravity Mediation}},  {\em JHEP} {\bf 09}
  (2016) 086, [\href{http://arxiv.org/abs/1608.06618}{{\tt arXiv:1608.06618}}].

\bibitem{Yanagida:2018eho}
T.~T. Yanagida, W.~Yin, and N.~Yokozaki, {\it {Flavor-Safe Light Squarks in
  Higgs-Anomaly Mediation}},  {\em JHEP} {\bf 04} (2018) 012,
  [\href{http://arxiv.org/abs/1801.05785}{{\tt arXiv:1801.05785}}].

\bibitem{hep-th/9810155}
L.~Randall and R.~Sundrum, {\it {Out of this world supersymmetry breaking}},
  {\em Nucl. Phys.} {\bf B557} (1999) 79--118,
  [\href{http://arxiv.org/abs/hep-th/9810155}{{\tt hep-th/9810155}}].

\bibitem{hep-ph/9409329}
T.~Yanagida, {\it {Naturally light Higgs doublets in the supersymmetric grand
  unified theories with dynamical symmetry breaking}},  {\em Phys. Lett.} {\bf
  B344} (1995) 211--216, [\href{http://arxiv.org/abs/hep-ph/9409329}{{\tt
  hep-ph/9409329}}].

\bibitem{hep-ph/9509201}
T.~Hotta, K.~I. Izawa, and T.~Yanagida, {\it {Dynamical models for light Higgs
  doublets in supersymmetric grand unified theories}},  {\em Phys. Rev.} {\bf
  D53} (1996) 3913--3919, [\href{http://arxiv.org/abs/hep-ph/9509201}{{\tt
  hep-ph/9509201}}].

%\cite{hep-ph/9607463}
\bibitem{hep-ph/9607463}
  N.~Arkani-Hamed, H.~C.~Cheng and T.~Moroi,
  {\it { Nonunified gaugino masses in supersymmetric missing partner models with hypercolor}},
  {\em Phys.\ Lett.\ B} {\bf 387} (1996) 529,
  %doi:10.1016/0370-2693(96)01089-1
  [\href{https://arxiv.org/abs/hep-ph/9607463} {{\tt hep-ph/9607463}}].
  %%CITATION = doi:10.1016/0370-2693(96)01089-1;%%
  %36 citations counted in INSPIRE as of 09 Jul 2019



\bibitem{1303.5830}
S.~Mohanty, S.~Rao, and D.~P. Roy, {\it {Reconciling the muon $g-2$ and dark
  matter relic density with the LHC results in nonuniversal gaugino mass
  models}},  {\em JHEP} {\bf 09} (2013) 027,
  [\href{http://arxiv.org/abs/1303.5830}{{\tt arXiv:1303.5830}}].

\bibitem{Iwamoto:2014ywa}
S.~Iwamoto, T.~T. Yanagida, and N.~Yokozaki, {\it {CP-safe gravity mediation
  and muon $g-2$}},  {\em PTEP} {\bf 2015} (2015) 073B01,
  [\href{http://arxiv.org/abs/1407.4226}{{\tt arXiv:1407.4226}}].

\bibitem{Izawa:2010ym}
K.~I. Izawa, T.~Kugo, and T.~T. Yanagida, {\it {Gravitational Supersymmetry
  Breaking}},  {\em Prog. Theor. Phys.} {\bf 125} (2011) 261--264,
  [\href{http://arxiv.org/abs/1008.4641}{{\tt arXiv:1008.4641}}].

\bibitem{hep-ph/0103067}
S.~P. Martin and J.~D. Wells, {\it {Muon Anomalous Magnetic Dipole Moment in
  Supersymmetric Theories}},  {\em Phys. Rev.} {\bf D64} (2001) 035003,
  [\href{http://arxiv.org/abs/hep-ph/0103067}{{\tt hep-ph/0103067}}].

\bibitem{hep-ph/9803384}
G.~Degrassi and G.~F. Giudice, {\it {QED logarithms in the electroweak
  corrections to the muon anomalous magnetic moment}},  {\em Phys. Rev.} {\bf
  D58} (1998) 053007, [\href{http://arxiv.org/abs/hep-ph/9803384}{{\tt
  hep-ph/9803384}}].

\bibitem{0808.1530}
S.~Marchetti, S.~Mertens, U.~Nierste, and D.~Stockinger, {\it
  {Tan(beta)-enhanced supersymmetric corrections to the anomalous magnetic
  moment of the muon}},  {\em Phys. Rev.} {\bf D79} (2009) 013010,
  [\href{http://arxiv.org/abs/0808.1530}{{\tt arXiv:0808.1530}}].

\bibitem{hep-ph/9912516}
M.~Carena, D.~Garcia, U.~Nierste, and C.~E.~M. Wagner, {\it {Effective
  Lagrangian for the $\bar{t} b H^{+}$ interaction in the MSSM and charged
  Higgs phenomenology}},  {\em Nucl. Phys.} {\bf B577} (2000) 88--120,
  [\href{http://arxiv.org/abs/hep-ph/9912516}{{\tt hep-ph/9912516}}].

\bibitem{1802.02995}
A.~Keshavarzi, D.~Nomura, and T.~Teubner, {\it {Muon $g-2$ and $\alpha(M_Z^2)$:
  a new data-based analysis}},  {\em Phys. Rev.} {\bf D97} (2018), no.~11
  114025, [\href{http://arxiv.org/abs/1802.02995}{{\tt arXiv:1802.02995}}].

\bibitem{hep-ph/0211331}
A.~Djouadi, J.-L. Kneur, and G.~Moultaka, {\it {SuSpect: A Fortran code for the
  supersymmetric and Higgs particle spectrum in the MSSM}},  {\em Comput. Phys.
  Commun.} {\bf 176} (2007) 426--455,
  [\href{http://arxiv.org/abs/hep-ph/0211331}{{\tt hep-ph/0211331}}].

\bibitem{1305.0237}
G.~Belanger, F.~Boudjema, A.~Pukhov, and A.~Semenov, {\it {micrOMEGAs\_3: A
  program for calculating dark matter observables}},  {\em Comput. Phys.
  Commun.} {\bf 185} (2014) 960--985,
  [\href{http://arxiv.org/abs/1305.0237}{{\tt arXiv:1305.0237}}].

\bibitem{1608.01880}
H.~Bahl and W.~Hollik, {\it {Precise prediction for the light MSSM Higgs boson
  mass combining effective field theory and fixed-order calculations}},  {\em
  Eur. Phys. J.} {\bf C76} (2016), no.~9 499,
  [\href{http://arxiv.org/abs/1608.01880}{{\tt arXiv:1608.01880}}].

\bibitem{1312.4937}
T.~Hahn, S.~Heinemeyer, W.~Hollik, H.~Rzehak, and G.~Weiglein, {\it
  {High-Precision Predictions for the Light CP -Even Higgs Boson Mass of the
  Minimal Supersymmetric Standard Model}},  {\em Phys. Rev. Lett.} {\bf 112}
  (2014), no.~14 141801, [\href{http://arxiv.org/abs/1312.4937}{{\tt
  arXiv:1312.4937}}].

\bibitem{hep-ph/9507294}
J.~A. Casas, A.~Lleyda, and C.~Munoz, {\it {Strong constraints on the parameter
  space of the MSSM from charge and color breaking minima}},  {\em Nucl. Phys.}
  {\bf B471} (1996) 3--58, [\href{http://arxiv.org/abs/hep-ph/9507294}{{\tt
  hep-ph/9507294}}].

\bibitem{hep-ph/9612464}
R.~Rattazzi and U.~Sarid, {\it {Large tan Beta in gauge mediated SUSY breaking
  models}},  {\em Nucl. Phys.} {\bf B501} (1997) 297--331,
  [\href{http://arxiv.org/abs/hep-ph/9612464}{{\tt hep-ph/9612464}}].

\bibitem{1011.0260}
J.~Hisano and S.~Sugiyama, {\it {Charge-breaking constraints on left-right
  mixing of stau's}},  {\em Phys. Lett.} {\bf B696} (2011) 92--96,
  [\href{http://arxiv.org/abs/1011.0260}{{\tt arXiv:1011.0260}}]. [Erratum:
  Phys. Lett.B719,472(2013)].

\bibitem{hep-ph/0206266}
T.~Nihei, L.~Roszkowski, and R.~Ruiz~de Austri, {\it {Exact cross-sections for
  the neutralino slepton coannihilation}},  {\em JHEP} {\bf 07} (2002) 024,
  [\href{http://arxiv.org/abs/hep-ph/0206266}{{\tt hep-ph/0206266}}].

\bibitem{1311.2162}
M.~Ibe, A.~Kamada, and S.~Matsumoto, {\it {Mixed (cold+warm) dark matter in the
  bino-wino coannihilation scenario}},  {\em Phys. Rev.} {\bf D89} (2014),
  no.~12 123506, [\href{http://arxiv.org/abs/1311.2162}{{\tt
  arXiv:1311.2162}}].

\bibitem{1403.0715}
K.~Harigaya, K.~Kaneta, and S.~Matsumoto, {\it {Gaugino coannihilations}},
  {\em Phys. Rev.} {\bf D89} (2014), no.~11 115021,
  [\href{http://arxiv.org/abs/1403.0715}{{\tt arXiv:1403.0715}}].

\bibitem{1803.02762}
{\bf ATLAS} Collaboration, M.~Aaboud et~al., {\it {Search for electroweak
  production of supersymmetric particles in final states with two or three
  leptons at $\sqrt{s}=13\,$TeV with the ATLAS detector}},
  \href{http://arxiv.org/abs/1803.02762}{{\tt arXiv:1803.02762}}.

\bibitem{1806.05264}
{\bf CMS} Collaboration, A.~M. Sirunyan et~al., {\it {Search for supersymmetric
  partners of electrons and muons in proton-proton collisions at $\sqrt{s}=$ 13
  TeV}},  {\em Submitted to: Phys. Lett.} (2018)
  [\href{http://arxiv.org/abs/1806.05264}{{\tt arXiv:1806.05264}}].

\bibitem{1712.08119}
{\bf ATLAS} Collaboration, M.~Aaboud et~al., {\it {Search for electroweak
  production of supersymmetric states in scenarios with compressed mass spectra
  at $\sqrt{s}=13$ TeV with the ATLAS detector}},  {\em Phys. Rev.} {\bf D97}
  (2018), no.~5 052010, [\href{http://arxiv.org/abs/1712.08119}{{\tt
  arXiv:1712.08119}}].

\bibitem{CMS-PAS-SUS-16-025}
{\bf CMS} Collaboration, {\it {Search for new physics in the compressed mass
  spectra scenario using events with two soft opposite-sign leptons and missing
  transverse momentum at $\sqrt{s}=13~\mathrm{TeV}$}},  Tech.
  Rep.\href{https://cds.cern.ch/record/2205866}{{ CMS-PAS-SUS-16-025}}, CERN,
  Geneva, 2016.

\bibitem{1309.3065}
M.~Endo, K.~Hamaguchi, T.~Kitahara, and T.~Yoshinaga, {\it {Probing Bino
  contribution to muon $g - 2$}},  {\em JHEP} {\bf 11} (2013) 013,
  [\href{http://arxiv.org/abs/1309.3065}{{\tt arXiv:1309.3065}}].

\bibitem{Yamaguchi:2016oqz}
M.~Yamaguchi and W.~Yin, {\it {A novel approach to finely tuned supersymmetric
  standard models: The case of the non-universal Higgs mass model}},  {\em
  PTEP} {\bf 2018} (2018), no.~2 023B06,
  [\href{http://arxiv.org/abs/1606.04953}{{\tt arXiv:1606.04953}}].

\bibitem{hep-ph/0007085}
A.~J. Buras, P.~Gambino, M.~Gorbahn, S.~Jager, and L.~Silvestrini, {\it
  {Universal unitarity triangle and physics beyond the standard model}},  {\em
  Phys. Lett.} {\bf B500} (2001) 161--167,
  [\href{http://arxiv.org/abs/hep-ph/0007085}{{\tt hep-ph/0007085}}].

\bibitem{LEP-slepton}
{\bf ALEPH, DELPHI, L3, OPAL Experiments} Collaboration, {\it {Combined LEP
  Selectron/Smuon/Stau Results}},  2004.
\newblock
  \href{http://lepsusy.web.cern.ch/lepsusy/www/sleptons_summer04/slep_final.html}{http://lepsusy.web.cern.ch/lepsusy/www/sleptons\_summer04/slep\_final.html}.

\end{thebibliography}\endgroup

\end{document}